# Photophoretic Movement of a Micron-Sized Light-Absorbing Capsule: Numerical Simulation

Yu. E. Geints[1],[*], E. K. Panina[1]

V.E. Zuev Institute of Atmospheric Optics, 1 Acad. Zuev square, Tomsk, 634021, Russia;
[*]Corresponding author e-mail: ygeints@iao.ru

**Abstract**

Multilayer microparticles with a liquid core and a polycomposite light-absorbing shell (microcapsules) are important components of modern bio- and medical technologies. Opening of the microcapsule shell and payload release can be realized by optical radiation. The photophoretic force is due to the radiation-stimulated thermal gradient and arises from the temperature inhomogeneity of the microparticle. Photophoretic forces, as well as radiation pressure forces, are inherently mechanical forces and can cause microcapsules to move during the opening cycle. We numerically simulate the microcapsule photophoretic motion when illuminated by an intense laser pulse. Numerical calculations of the temperature field in a spherical microcapsule are carried out using the finite element method, taking into account the auxiliary nanoparticles, which are randomly distributed around the capsule and serve to enhance the heating of the capsule under short pulse exposure. The spatial distribution of the absorbed optical power as well as the temporal dynamics of microcapsule heating depending of its size are investigated in detail. We show, for the first time to our knowledge, that under the action of photophoretic gradient, the microcapsule can move along the laser incidence direction both forward and backward at the distance of several tens of nanometers depending on the particle size and conditions of optical absorption.

## 1. Introduction

In the past few decades, the technology of microencapsulation of various substances has been increasingly developed for use in numerous branches of industry, science and technology [1-3]. Microencapsulation is a process in which nanodoses and nanoparticles of liquid or solid substances are surrounded by a continuous film of polymeric material, which in turn isolates them from the external environment and is mainly used for protection as well as controlled payload release [4]. Generally, capsules are two-component microspheres consisting of a thin multilayer polymer shell and a liquid core with active substance. An undoubted breakthrough of modern engineering technologies was the creation of multilayer composite capsules by the layer-by-layer (LbL) shell assembly [5], which enables expanding the area of their possible practical application. As assumed, the microencapsulation can be used to solve a number of important problems in pharmacology, such as extending the expiration dates of rapidly deteriorating drugs, masking their properties, separate reactive drugs combined in a common dosage form, and reducing the drug reactivity [6].

To impart the multifunctionality to a microcapsule, its shell is usually assembled from several heterogeneous organic/inorganic layers that respond to different external stimuli. For example, the addition of iron oxide ($Fe_3O_4$) nanoparticles (NPs) imparts magnetic properties to the microcapsule [7], the introduction of polycrystalline titanium oxide ($TiO_2$) allows the creation of a transport container with improved strength properties, reduced wall permeability, susceptibility to ultrasound as well as UV irradiation [8]. The sensitivity of the capsule shell to an optical radiation is usually provided by the introduction of substances that actively absorb light within a desired spectral range. Thus, in the visible and near-infrared (IR) spectral regions, shell doping with noble metal NPs (silver, gold, gold sulfide) [9], or addition of dye molecules [10] are usually exploited.

When illuminated by an optical radiation, a small suspended particle experiences a light pressure force and a photophoretic force, which arises due to the interaction of molecules of the surrounding medium (gas, liquid) with the surface of the particle that is heated by light unevenly. Ambient molecules reflected from the hot side of the particle after inelastic collisions move faster than those reflected from the colder side, and as a result the particle acquires net mechanical momentum of motion.

The multifunctionality of a microcapsule, as a transport microcontainer, also implies the possibility of its active targeted transportation to the desired point of a biological or other substance. As a rule, the most common method of the so-called, system transport administration [11], of a microencapsulated cargo is the use of special nanoscale carriers in the form of metal NPs controlled by a magnetic field [12], or micro- and nanoswimmers activated by the phoretic forces of a chemical or magnetic gradient [13]. At the same time, it was shown in [14] that specially engineered micron-sized Janus particles with spatially inhomogeneous chemo-thermal properties without additional carriers can also move self-phoretically and orient themselves due to catalytic reactions in the presence of an electromagnetic field. The self-movement of a colloidal Janus-particle with a semi-metallic coating under the action of laser irradiation caused by the photophoresis was experimentally studied in [15].

Since the photophoresis arises from the exchange of momentum between the particle and surrounding molecules this distinguishes it from optical radiation pressure forces, where momentum transfer occurs between the particle and photons. As a consequence, the photophoretic force is usually many times stronger than the radiation pressure force. The fundamental condition for the manifestation of photophoresis forces is the presence of optical absorption in the substance of the microparticle leading to the appearance of a temperature gradient and a change in the kinetics of the medium near the illuminated and shadow parts of the particle.

In present work, we theoretically address the motion in a light beam of a micron spherical core-shell particle that mimics a microcapsule with an optically absorbing shell and a water core which possible contains some payload. We numerically simulate the temporal dynamics of the spatial distribution of capsule temperature and calculate the mechanical forces arising from the thermal gradient phoresis and direct light pressure. We show that the competition between these forces can lead to a change in the direction of motion of microcapsule with respect to the direction of laser incidence.

## 2. Photophoresis of an absorbing spherical microparticle

Generally, two types of physical photophoresis are distinguished, usually denoted as $F_{\Delta T}$ and $F_{\Delta \alpha}$ [16], which are due to the different type of momentum exchange between molecules and the particle. The force due to the radiation-stimulated thermal gradient ($F_{\Delta T}$) arises from the inhomogeneity of the particle temperature. The molecules of the medium surrounding the particle leave its hotter surface with more energy on average than molecules leaving the colder part, thus creating an effective mechanical force directed from the hot to the cold side of the particle. Meanwhile, under the action of the phoretic force $F_{\Delta T}$ the particle moves along the axis of laser radiation action (both forward and backward [17]), which with a proper configuration of the microsystem can balance the force of gravity and lead to levitation of the particle with respect to Earth's gravitation [18-20].

The second type of photophoresis ($F_{\Delta \alpha}$) is related to the kinetics of momentum exchange between the medium molecules and the particle regulated by the accommodation coefficient. The thermal accommodation coefficient α represents the probability that a medium molecule colliding with a particle will come into thermal equilibrium with its surface. Values of the accommodation coefficient vary widely, from approximately 0.95 to 0.5, and depend on the structure and morphology of the particle itself and the type of environment [21]. In contrast to the thermal gradient photophoresis $F_{\Delta T}$, the "momentum" force $F_{\Delta \alpha}$ is related to the coordinate system of a particular particle (not optical wave) and acts also on isothermally heated particles as long as there is a difference in accommodation coefficients between its parts. Typically, the Δα gradient occurs in irregularly shaped particles (planetary dust, biological objects), and/or with inhomogeneous bulk absorption distribution (Janus particles) [23, 24] and leads to spatial reorientation of the particle under the action of the photophoretic torque. In the following, we consider only the perfect homogeneous spherical microparticles, so we will neglect the accommodative component of photophoresis and consider only the forces associated with the temperature gradient $\Delta T$ of the particle surface.

In the continuous medium approximation, which is valid at small Knudsen numbers $Kn = l/R \ll 1$, where $l$ is the average path length of the medium molecule and $R$ is the radius of the particle, the following expression can be used for the photophoresis force $F_{\Delta T}$ [24]:

$$F_{\Delta T} = D\left(p^*/p\right) R \Delta T. \tag{1}$$

Here, the coefficient $D$ is determined only by the properties of the environment surrounding the particle, $p$ is environmental pressure, and $p^*$ is some reference pressure depending among other things on the particle radius $R$ [16]:

$$D = \frac{\pi}{2}\sqrt{\frac{\pi \xi}{3}} \frac{\bar{c}\eta}{T}, \qquad p^* = \frac{3DT}{\pi R} \tag{2}$$

In Eq. (2), $\xi$ is the thermal slip coefficient of medium molecules ($\xi \approx 1$ [24, 25]), $\bar{c}$ stays for the average velocity of medium molecules (for water we assume $\bar{c} \approx 650$ m/s) with viscosity $\eta$, and temperature $T$. Carrying out simple calculations for typical values of water parameters at 20°C ($p = 1.13 \cdot 10^5$ Pa), we obtain a convenient formula for calculating the photophoresis force:

$$F_{\Delta T}[\text{nN}] = 26.4 \cdot \left(\Delta T/1[\text{K}]\right), \tag{3}$$

where the force is expressed in nanonewtons (nN), while the parameter $\Delta T$ defines the average temperature difference between the illuminated and shadow parts of the particle. It can be seen that depending on the sign of this value, the photophoresis can be omnidirectional, i.e. both along and against the action of optical radiation. Despite the rather simple form of Eq. (3), the main difficulty in its use in practical calculations of photophoresis is the calculation of the value $\Delta T$ since it requires solving the complete problem of particle heating by optical radiation.

Indeed, according to the definition the temperature difference is expressed through the following integrals:

$$\Delta T = T_1 - T_2 = \frac{1}{S_1}\oint_{S_1} T(s)\,ds - \frac{1}{S_2}\oint_{S_2} T(s)\,ds, \tag{4}$$

where the integration is performed over the surfaces of the illuminated ($S_1$) and shadow ($S_2$) particle hemispheres. The surface temperature distribution is found from the solution of the heat transfer problem inside and outside the particle.

The corresponding thermophysical problem is usually considered in the isobaric approximation, when the inequality $t^* \gg R/c_s$ holds true. Here, $t^*$ is the characteristic time scale of the problem, e.g., laser pulse duration $t_p$, $R$ is the equivalent particle size, and $c_s$ is sound velocity in the medium. In the conditions when the pressure equalization in the cold and heated regions of the condensed medium is sufficiently fast (on the scale of an optical pulse), a single equation for the nonstationary particle temperature can be used instead of the full closed system of Navier-Stokes equations [Carslaw59]:

$$\rho\, C_p \frac{\partial T(\mathbf{r},t)}{\partial t} = \lambda_T \nabla^2 T(\mathbf{r},t) + Q(\mathbf{r},t). \tag{5}$$

Here $\nabla^2$ denotes the Laplace operator in 3D-space, $\rho$, $C_p$, $\lambda_T$ are density, specific (isobaric) heat and heat conductivity of medium, respectively, $Q(\mathbf{r},t)$ is volumetric density of heat sources due to optical absorption by a particle. All thermodynamic parameters are considered dependent on the spatial coordinate and medium temperature. In the following, the medium temperature $T$ is treated as the difference $\Delta T$ relative to the initial value $T_0 = 293$ K. By using Eq. (5), it is assumed the absence of convective heat fluxes between differently heated parts of medium, which is normal for considering heat transfer on the microscales. Besides, we also neglect any advection flows around the particle. The outer boundaries of the computational domain are set as isothermal, e.g., there is no heat transfer outside the domain and the following condition is satisfied: $\nabla T|_{\mathbf{r}=\mathbf{R}} = 0$.

In Eq. (5), the density of thermal sources $Q(\mathbf{r},t)$ depends on the spatial and temporal profiles of the optical intensity $I(\mathbf{r},t)$ in the microcapsule: $Q(\mathbf{r},t) = \alpha(\mathbf{r}) I(\mathbf{r},t)$, where $\alpha = 2\pi\varepsilon''/\lambda$ is the bulk absorption coefficient, $\varepsilon''$ is imaginary part of the dielectric permittivity of particle substance, and $\lambda$ denotes the wavelength of the optical radiation. Generally, to obtain the optical field inside a multilayer particle illuminated with a short laser pulse one has to solve the nonstationary optical scattering problem based on the full-wave equation for the electromagnetic field. However, this problem can be significantly simplified if we do not consider transient processes of establishing the stationary optical field distribution in the vicinity of the particle, but rather consider the field to be quasi-stationary with a spatial profile, whose amplitude follows the temporal profile of the laser pulse.

Then, by the moment when the thermodynamic equilibrium in the medium is established, the optical field at every time instance $t$ obeys the homogeneous Helmholtz equation:

$$\nabla \times \nabla \times \mathbf{E}(\mathbf{r};t) - k^2 \varepsilon(\mathbf{r}) \mathbf{E}(\mathbf{r};t) = 0. \tag{6}$$

Here, $\varepsilon = \varepsilon' + j\varepsilon''$ is the complex dielectric permittivity of medium, $k = 2\pi/\lambda$ is the wave number. The electric field $\mathbf{E}$ is understood as a vector with components along each of the coordinate axes and initial amplitude $E_0(t)$. Worthwhile noting, for multilayer spherical or cylindrical particles the analytical solution to Eq. (6) is known in the form of infinite series on multipoles (spherical vector harmonics), called in the literature the "Mie series" [26]. However, there is no such analytical solution for a microsphere surrounded by an auxiliary NP cloud, so next we use numerical solution to Eq. (6).

### 3. Numerical model, results and discussion

For the numerical integration of Eqs. (5)-(6) we employ COMSOL Multiphysics software, which implements the finite element method (FEM) for the solution of differential equations of physics. Structurally, the microcapsule is represented by a two-layered core-shell microsphere with a non-absorbing aqueous core and a strongly light-absorbing solid-phase shell (Fig. 1a). A silicone doped with gold nanorods is assumed as the shell material, which provides the desired optical absorption spectrum. According to Ref. [27], the gold nanorods demonstrate enhanced optical absorption in the near- IR region of the spectrum, approximately from 700 nm to 900 nm (Fig. 1b). In the simulations, the capsule radius $R_c$ is fixed to 500 nm with a shell thickness $h = 65$ nm. The gold nanorods inside the capsule shell produce an absorption peak at $\lambda = 755$ nm corresponding to the excitation of the longitudinal plasmonic mode [27]. Due to the chaotic spatial orientation of the nanorods with respect to the plane of optical wave polarization, the spectral contour of the plasmon resonance broadens and evolves to a smooth absorption hump within the range from 730 nm to 790 nm.

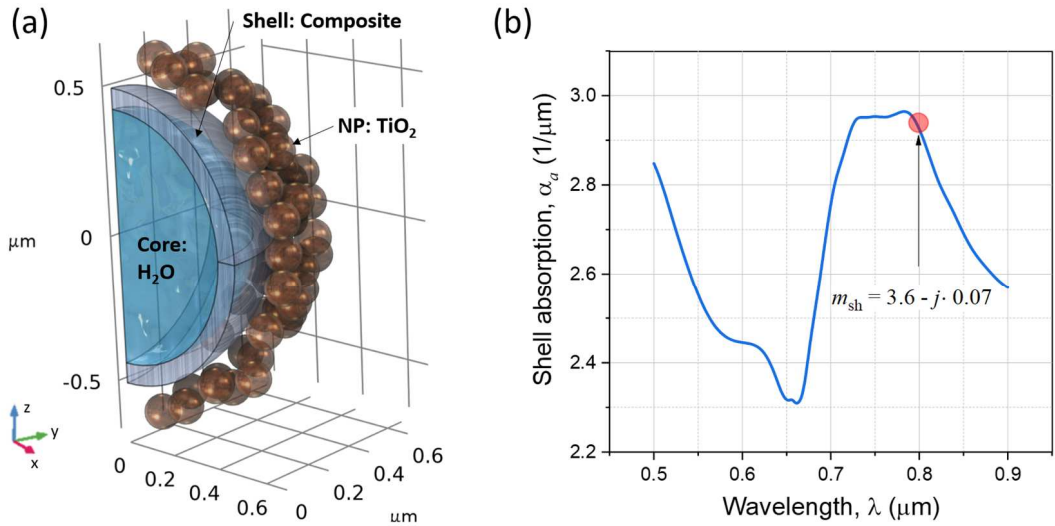

Fig. 1. (a) 3D-model of a microcapsule with an absorbing nanocomposite shell and an aqueous core surrounded by an ensemble of aux-NPs (TiO$_2$). (b) Optical absorption spectrum of microcapsule with volume fraction of gold nanorods in shell, $\delta_{Au} = 0.1$. The absorption peak at the wavelength of 800 nm is marked by an arrow.

Further, for the sake of calculation simplicity, the microstructural composition of the capsule shell is not accounted, but certain homogeneous medium with an effective complex refractive index $m_{sh}$ is considered, which is calculated through the Bruggeman effective medium model at a volume fraction of nanorods in the shell, $\delta_{Au} = 0.1$. Thus, for chosen laser wavelength $\lambda = 800$ nm (the fundamental harmonic of a Ti:Sapphire-laser), the effective shell refractive index has the value, $m_{sh} = 3.6 - j \cdot 0.07$. Microcapsule core and the surroundings are considered as water with refractive index $m_0 = 1.33$ and zero absorption of near-IR radiation.

Naturally, one of the best ways to improve the microcapsule perforation by optical radiation is increasing the optical field concentration on its shell. To this end, an auxiliary aerosol agent consisting of dielectric or metallic NPs randomly distributed near the microcapsule was proposed recently [28]. This leads to substantial growth of capsule optical absorption that is related both to the focusing of the optical field by a transparent nanoparticle on the nanoscale with the formation of a "photonic nanojet" and to the collective oscillations of electrons on the surface of plasmonic nanoparticles.

In accordance with these considerations, in the present study we place an ensemble of auxiliary NPs around the microcapsule, whose (NPs) spatial position is calculated using the original algorithm [28]. In this case, an array of the coordinates of auxiliary NP centers is generated inside a given globular layer encompassing the capsule. NPs are distributed randomly under the condition of neighboring particles non-overlapping. As the material of auxiliary NPs, a biocompatible substance such as titanium dioxide ($TiO_2$) is specified.

In turn, to set the thermophysical properties of the composite metal-dielectric microcapsule shell, we use the COMSOL module "Porous Medium", which takes into account heat transfer in a two-component medium consisting of a porous matrix (silicone) and a heat-transfer dope (Au) with certain volume fraction $\delta_{Au}$. This resembles the transition from a two-phase to a homogeneous medium with effective parameters calculated by the following models [29]:

$$(\rho)_e = \delta_{Au} (\rho)_d + (1-\delta_{Au})(\rho)_m, \qquad (7a)$$

$$(\rho C_p)_e = \delta_{Au} (\rho C_p)_d + (1-\delta_{Au})(\rho C_p)_m, \qquad (7b)$$

$$(1/\lambda_T)_e = \delta_{Au} (1/\lambda_T)_d + (1-\delta_{Au})(1/\lambda_T)_m. \qquad (7c)$$

Here, the subscripts "e", "d", "m" denote the values of the effective (volumetric) parameter and the corresponding values for the dope (gold) and matrix (silicone). The relevant thermophysical parameters of the microcapsule shell calculated from Eqs. (7a)-(7c) are given at $\delta_{Au} = 0.1$ and are as follows: $\rho = 3.07$ g/cm$^3$, $C_p = 1.63$ kJ/(kg·K), $\lambda_T = 0.29$ W/(m·K).

A Gaussian pulse of circularly polarized optical radiation with typical parameters of $\lambda = 800$ nm and duration $t_p = 1$ ps, is directed to the microcapsule. The optical radiation is focused into a 8 μm diameter waist with the total pulse energy 2 nJ that gives peak pulse intensity $I_0$ of about 1 GW/cm$^2$.

Figs. 2(a-d) show the distribution of normalized optical intensity $I = |\mathbf{E}|^2 / E_0^2$ ($E_0$ is initial electric field amplitude) and volume density of heat release sources $Q$ in the principal cross-section of a spherical microcapsule to the end of laser pulse. In addition, for clarity in (a) and (b), the streamlines of the Poynting vector $\mathbf{S} = (c/8\pi) \operatorname{Re}[\mathbf{E} \times \mathbf{H}^*]$ ($\mathbf{H}$ is the magnetic field) are plotted showing the direction of optical energy fluxes inside the capsule.

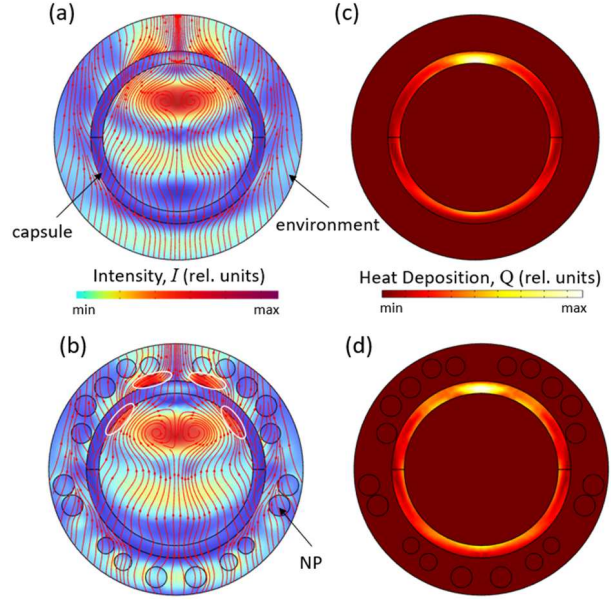

Fig. 2. (a,b) Spatial distribution of optical intensity $I$ and (c,d) power of heat release $Q$ near the microcapsule without (a,c) and with (b,d) auxiliary NPs. Colored streamlines show the energy flow structure (Poynting vector **S**) in the capsule. In (b), red ellipses highlight the regions of increased optical field concentration.

Two situations are considered: (a) when the microcapsule is in "clean" conditions, i.e., in a homogeneous medium (water) without additional nanoscatterers, and (b) when the capsule is surrounded by an ensemble of dielectric NPs fabricated of titanium dioxide with a diameter of 120 nm and refractive index $m_{NP} = 2$. In both cases, due to focusing of light by the capsule in its shadow hemisphere a maximum of field intensity is formed, which leads to a corresponding increase of heat generation in this region. Placing the auxiliary nano-aerosol in the vicinity of the capsule changes the structure of the energy flows, forcing the optical field to be additionally concentrated in the regions of microfocusing by the NPs. This in turn enhances the absorption of the optical radiation by the capsule shell in the areas where the multiple foci from the NP layer is formed (shown as red ovals in Fig. 2b), and, as a result, can affect the temperature distribution in the capsule and its photophoretic ability.

An example of the spatial distribution of optical intensity and temperature in different regions of the capsule and surrounding dispersed medium is shown in Figs. 3(a) and (b). The laser pulse parameters are same as in Fig. 2. The temperature distribution is calculated at the time instant $t = 75$ ps. As seen, with the chosen parameters of the photonic microstructure, the maximum of the field intensity is located inside the capsule core, which is considered to be non-absorbing. However, due to the lensing effect, the optical field intensity rises on the surface of the shadow hemisphere, in the absorbing capsule shell. In this case, the NPs which surround the capsule scatter the optical wave and create additional areas of light microfocusing on the capsule shell. As a result, the shadow region of the microcapsule becomes more heated than its illuminated part.

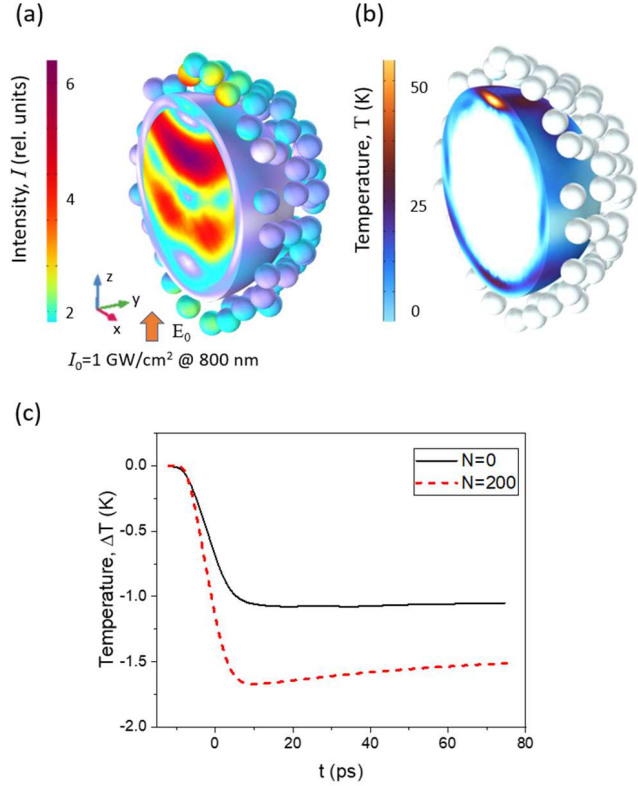

Fig. 3. 3D-distribution of (a) optical intensity $I$ and (b) temperature $T$ in the vicinity of microcapsule with auxiliary NPs. The laser pulse parameters are shown in the figure. (c) Temporal dynamics of the averaged temperature difference $\Delta T$ between lower and upper capsule hemispheres at different number of auxiliary NPs.

This can be clearly seen in Fig. 3(c), where the difference of averaged temperature $\Delta T$ calculated for the lower and upper hemispheres of the capsule is plotted as a function of time and the number $N$ of auxiliary NPs. The time scale in this plot is specially shifted so the pulse center corresponds to $t = 0$. The dependences presented demonstrate heating of the capsule during the pulse and following gradual cooling of the non-uniformly heated shell due to heat diffusion into the cold capsule core and the environment during the characteristic time-scale of thermal relaxation $t_T = h^2/4\chi \approx 10$ ns, where $\chi = \lambda_T/\rho C_p$ is the coefficient of thermal conductivity of the shell. Negative $\Delta T$ values show that for a micron-sized capsule negative photophoresis is realized and under the action of thermophoretic forces the particle tends to move towards optical radiation.

Worth recalling, the particle illuminated by an intense optical radiation is also subjected to the optical pressure force $\mathbf{F}_{opt}$ arising due to compensation of the net momentum of photons during their scattering and absorption on particle surface. In the absence of a longitudinal gradient of the background optical field, the light pressure force acts along the wave vector of optical radiation and tends to push the microparticle along the direction of light propagation. Thus, to determine the resultant displacement of heated microcapsule one should consider the resultant (net) optical force:

$\mathbf{F}_{net} = \mathbf{F}_{\Delta T} + \mathbf{F}_{opt}$.

We obtain the optical pressure forces through the full-wave calculations of the Maxwell stress tensor $\mathbf{T}_S$ associated with the Poynting flux $\mathbf{S}$. In isotropic medium, the following expression for the integral optical force $\mathbf{F}_{opt}$ [30] is used for temporally averaged fields:

$$\mathbf{F}_{opt} = -\oiint_S (\mathbf{T}_S \cdot \mathbf{e}_n) ds \tag{8}$$

$$T_{S_{ij}} = \frac{1}{2}\delta_{ij}\left(\varepsilon|\mathbf{E}|^2 + |\mathbf{H}|^2\right) - \left(\varepsilon E_i E_j + H_i H_j\right) \tag{9}$$

Here, the integral is taken along the capsule surface with the external normal $\mathbf{e}_n$, while $T_{S_{ij}}$ denote the (*ij*)-components of the stress tensor. The medium and the particle are assumed to be nonmagnetic.

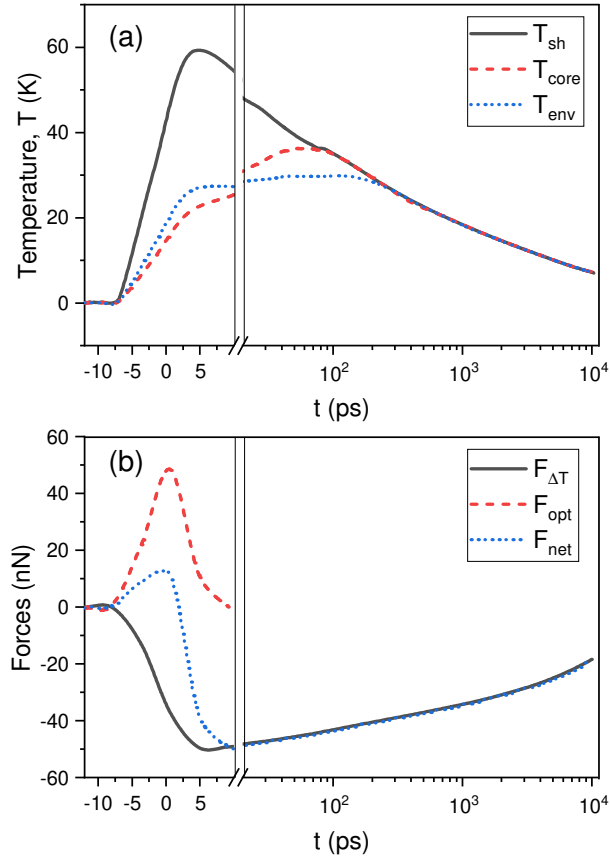

Fig. 4. (a) Temporal dynamics of maximal temperature in shell $T_{sh}$, core $T_{core}$ and surrounding fluid $T_{env}$. (b) Amplitude of heat-induced photophoretic force $F_{\Delta T}$, optical pressure $F_{opt}$ and resultant force $F_{net}$ for a capsule with $R_c = 500$ nm exposed to a 1 ps pulse with the intensity of 1 GW/cm$^2$.

The temporal dependence of the maximum values of the temperature increment in different regions of the capsule and the environment, as well as the dynamics of the optical field forces are shown in Figs. 4(a) and (b). Here, similar to Fig. 3(c), two physical processes corresponding to different time scales are clearly visible. The first one is the laser heating of the core-shell particle during the laser pulse acting interval of approximately 10 ps (~ 10·$t_p$ defined on the level $I/I_0 = 1\%$). The second process describes the cooling of a non-uniformly heated particle due to heat diffusion into the cold capsule core and the environment.

As seen, at the first stage (heating) when the laser radiation is still acting, the temperature of capsule shell $T_{sh}$ considerably exceeds the core temperature $T_{core}$ and the temperature of the surrounding liquid $T_{env}$. Meanwhile, at this stage the optical pressure $F_{opt}$ prevails over the photophoresis $F_{\Delta T}$ and the net optical force $F_{net}$ is directed along the incident wave vector. Contrarily, after the end of the laser heating stage when the thermal relaxation of capsule takes place, the force of direct light pressure disappears, the temperature in all capsule regions gradually equalizes, and the resultant of forces **F**$_{net}$ changes its direction to the opposite pulling the microparticle in the opposite direction of the light beam incidence.

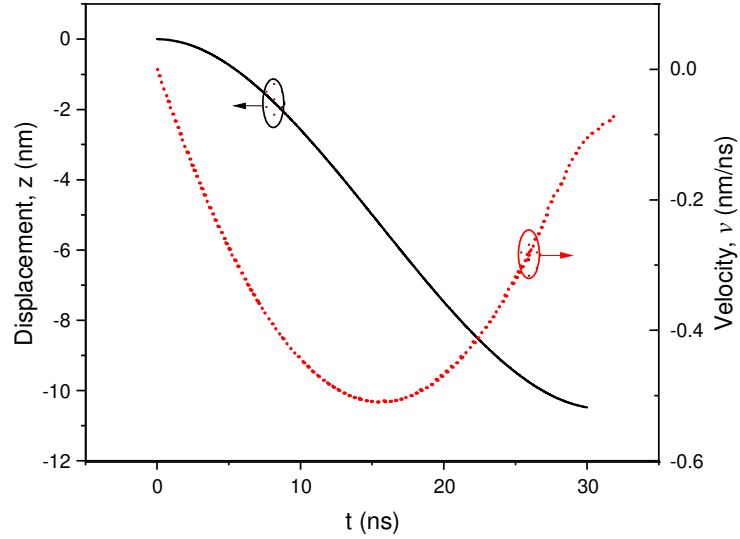

Fig. 5. Dynamics of velocity $v$ and displacement $z$ of a 500 nm microcapsule when exposed to an optical pulse with an energy of 2 nJ.

To calculate the magnitude of possible displacement of the microcapsule under the action of optical forces we use the equation of motion of a tiny solid particle with mass $m$ exposed to an external force. Following the Newton's law of motion, we have:

$$m\frac{d^2z}{dt^2} = F_{net} - \gamma\frac{dz}{dt} \qquad (10)$$

Here, $F_{net}$ is the resultant of optical forces acting along z-axis, $\gamma = 6\pi\eta R_c$ is the Stokes friction for a spherical particle with radius $R_c$ moving in a homogeneous medium with dynamic viscosity $\eta$ (for water, $\eta = 8.9 \cdot 10^{-4}$ Pa·s). By writing Eq. (10), we consider that the Earth gravitation is balanced by the particle buoyancy in water. The velocity $v = dz/dt$ and travel distance $z$ of a 500 nm-radius capsule exposed to a 1 ps laser pulse with the peak intensity of 1 GW/cm$^2$ are plotted in Fig. 5.

Clearly, the inertia of microparticle, $m(dv/dt)$, prevents its instantaneous displacement during a short time interval of the laser pulse action. Therefore, the action of light pressure forces practically does not affect capsule displacement and the whole particle dynamics is determined by more "long-playing" photophoretic forces, which become increasing in the time interval $t \leq 10$ ns in accordance with the temperature dynamics (Fig. 4b). As seen in Fig. 5, the maximum distance that a microcapsule can move under the action of a single pulse is of the order of 10 nm, and this distance is accumulated within about 30 ns. This obtained capsule displacement although small needs to be considered, since common ultrashort laser source operates in the pulse-periodic mode with pulse repetition rates typically of the tens of kHz. Therefore, during the interpulse period the capsule has time to accelerate and stop its motion. Then, the next laser pulse in a train will again initiate the capsule displacement and so on. As a result, the final distance of a microcapsule displacement, e.g., for one second of laser irradiation can be already about a millimeter. Obviously, this fact should be taken into account when designing laser-stimulated transportation microcontainers.

Importantly, when decreasing the capsule size its photophoretic displacement increases; an increase in the capsule radius, on the contrary, leads to the predominance of optical pressure over phoresis and the displacement distance decreases. Thus, e.g., under the same conditions of optical irradiation, a capsule with $R_c = 300$ nm can move towards the radiation by 64 nm in a time interval of about 15 ns. At the same time, a capsule with a radius of 750 nm can displace only for 3 nm but along the pulse propagation direction.

Worth noting, such considerable capsule displacements can cause its elastic collisions with surrounding NPs and the rearrangement of the nano-aerosol cloud. This can change the absorption efficiency of optical radiation by the capsule and affect its phoretic motion. However, in this work we do not address this rather complicated problem of many-body interaction and consider that a heavy microcapsule moves adiabatically together with its nearly weightless surroundings and preserved the NP spatial arrangement.

## Conclusion

In conclusion, we present the numerical simulation of the photophoresis of an absorbing core-shell spherical micron particle - a microcapsule, illuminated by an intense ultrashort laser pulse. The photophoretic force arises due to the light-stimulated thermal gradient between the illuminated and shadow parts of the microcapsule. We consider the situation when the microcapsule is additionally surrounded by an ensemble of transparent auxiliary nanoparticles, which help to enhance the optical absorption of the mother capsule. We show that under the action of the net optical force, a 500 nm microcapsule moves along the laser direction both forward and backward depending on the conditions of optical absorption and the ratio of the amplitudes of the light pressure and photophoretic force. The maximum distance of capsule movement under the action of a single laser pulse depends on the size of the microparticle and can reach tens of nanometers for a pulse with picosecond duration and peak intensity of 1 GW/cm$^2$.

**Funding**. Russian Science Foundation (#23-21-00018).

**Disclosures**. The author declares no conflicts of interest.

**Data availability**. Data underlying the results presented in this paper may be obtained from the authors upon reasonable request.